\documentclass[twocolumn]{aastex631}

\begin{document}

\title{Photometry of Saturated Stars with Neural Networks}

\shortauthors{Winecki \& Kochanek}

\author[0009-0009-8632-3102]{Dominik Winecki}
\affiliation{Department of Computer Science and Engineering,
The Ohio State University,
Columbus, OH 43210, USA}

\author[0000-0001-6017-2961]{Christopher S. Kochanek}\thanks{Email: kochanek.1@osu.edu}
\affiliation{Department of Astronomy, The Ohio State University, 140 W. 18th Ave., Columbus, OH 43210, USA}
\affiliation{Center for Cosmology and AstroParticle Physics, The Ohio State University, 191 West Woodruff Ave., Columbus, OH 43210, USA}






\begin{abstract}
We use a multilevel perceptron (MLP) neural network to obtain photometry of saturated
stars in the All-Sky Automated Survey for Supernovae (ASAS-SN).  The MLP
can obtain fairly unbiased photometry for stars from $g\simeq 4$ to $14$~mag,
particularly compared to the dispersion (15\%-85\% $1\sigma$ range around the median) of $0.12$~mag
for saturated ($g<11.5$~mag) stars.  More importantly, the light
curve of a non-variable saturated star has a median dispersion of only $0.037$~mag.  The MLP
light curves are, in many cases, spectacularly better than those provided by 
the standard ASAS-SN pipelines. While the network was trained on g band data 
from only one of ASAS-SN's 20 cameras, initial experiments suggest that
it can be used for any camera and the older ASAS-SN V band data as well.  The dominant problems seem to be associated
with correctable issues in the ASAS-SN data reduction pipeline for saturated stars
more than the MLP itself. The method is publicly available as a light curve
option on ASAS-SN Sky Patrol v1.0.
\end{abstract}



\section{Introduction} \label{sec:intro}

Projects such as the All-Sky Automated Survey for Supernovae (ASAS-SN, \citealt{Shappee2014},\citealt{Kochanek2017}, \citealt{Hart2023}), the Asteroid Terrestrial-impact Last Alert System (ATLAS, \citealt{Tonry2018}) and the Zwicky Transient Facility (ZTF, \citealt{Bellm2019}) currently monitor
all (ASAS-SN and ATLAS) or large fractions (ZTF) of the visible sky and
provide public access to their data.  For ASAS-SN, this consists of 
its catalog of variable stars (\citealt{Jayasinghe2019}), ASAS-SN Sky Patrol (\citealt{Kochanek2017}),
which will provide an uncensored light curve for any user-requested
sky coordinate, and ASAS-SN Sky Patrol v2.0 (SP2, \citealt{Hart2023}),
which provides continuously updated light curves of roughly 100
million sources from asteroids to stars and quasars.  In the near future,
the Vera Rubin Observatory (e.g., \citealt{Ivezic2019}) will provide ZTF-like coverage and
cadence for far fainter sources.  

Professional astronomy has lost, however, the 
ability to monitor the bright sky -- the need for both funding and efficient 
operations drive projects towards fainter and more numerous sources. Yet
these very bright sources are the ones which will have the best stellar
spectroscopic observations from projects like SDSS/APOGEE (e.g., \citealt{Abdurrouf2022}) or 
Gaia (e.g., \citealt{Gaia2023})
including either indications of binarity or full orbital solutions.

The problem is that charge coupled device (CCD) detectors have a finite dynamic 
range above which a pixel is saturated. A typical dark time
All-Sky Automated Survey for Supernovae (ASAS-SN)
image has a sky background of $\sim 200$ counts and saturates at
$\sim 60,000$ counts. The PSF has a FWHM of roughly 2 pixels
(\citealt{Shappee2014}, \citealt{Kochanek2017}), leading to a
point source dynamic range of roughly 7~magnitudes from
$g \sim 11.5$ to $g\sim 18.5$~mag.  This leaves some $\sim 10^6$
stars that are saturated in ASAS-SN observations. The excess
charge of the saturated pixels then bleeds into 
pixels along the read direction of the detector.

ASAS-SN inherited from the original All-Sky Automated Survey
(ASAS, \citealt{Pojmanski2002}) one method of trying to
obtain photometry of saturated stars. The pipeline tries
to identify the flux in the bleed trails of the saturated stars
and then adds a Gaussian with that flux at the star's location.
The bleed trails are then replaced by linear
interpolation of the adjacent, unsaturated pixels. This
approach has also been applied to Swift/UVOT data by
\cite{Page2013}.  As shown by some of the examples in
\cite{Kochanek2017}, this procedure frequently works
surprisingly well.  An alternative approach is to model
the unsaturated wings of the PSF (e.g., \citealt{Su2022}, \citealt{Zhou2023}).

Here, we experiment with a new approach, namely, training
a multilevel perceptron (MLP) neural network to do photometry of
saturated stars.  Convolutional
neural networks (CNNs) have been used successfully for
a range of astronomical problems, such as star–galaxy
classification (\citealt{kim_stargalaxy_2017}), transient
classification (\citealt{qu2022convolutional}) and
photometric redshifts (\citealt{pasquet_photometric_2019}).
There have been some applications of CNNs to photometry 
(\citealt{yang_star_2023,yuan_red_2023}) but not to the
specific problem of saturated star photometry. Our 
data consists of small, non-rectangular images centered
on each target star, and the relevance of each input pixel
is highly dependent on its position, so a MLP seemed a
better choice for this purpose than a CNN.
The theory is that a network can simply learn
the sensor-specific behavior for stars of different levels of saturation
and then predict the true brightness. 
In Section \ref{sec:data} we describe the construction of the training set,
and in Section \ref{sec:methods} we describe the model and the training
process.  We present the results in Section \ref{sec:results} with comparisons
to the results produced with aperture photometry on subtracted images by
ASAS-SN Sky Patrol v2.0 (\citealt{Hart2023}). We discuss the
results, known problems, potential solutions and possible future
improvements in Section \ref{sec:conclusion}.

\section{Training Data} \label{sec:data}

We selected stars from Gaia DR3 (\citealt{Gaia2016},
\citealt{Gaia2023}) that were observed by the ``bi'' camera of the ASAS-SN
Bohdan Paczynski mount at CTIO in 
Chile.  We selected a random 40,000 non-variable stars (based on
Gaia DR3, \citealt{Gaiavar2023}) over the sky per one magnitude
wide bin starting at $G=2$~mag. This provides a training 
set roughly uniformly distributed in magnitude.
Obviously, the very 
brightest bins have fewer than 40,000 stars and so we 
were simply including all available stars.  We kept
stars observed by bi, with defined $G$, $B_P$ and $R_P$
magnitudes.

Each target then has 100s of epochs of ASAS-SN observations, so
we randomly selected $\sim 4$ images per target and 
extracted a 21 pixel square ``postage stamp'' image of
the target.  We used single exposures interpolated to
the astrometric frame of the reference image.  Only
images flagged as having been taken in good conditions
were used and the postage stamp edges had to be at least
50 pixels from the detector edges.  We generated
$\sim 332$~thousand postage stamps spanning $G=3$
to $15$~mag.

We predict the g band magnitudes from the Gaia DR3
magnitudes because the Gaia magnitudes homogeneously
span the full range we consider.
We first applied
the Gaia EDR3 saturation corrections from \citealt{Riello2021}
(these affect $G<8$ for G, $G<3.94$ for $B_P$ and $R_P<3.45$ for $R_P$) 
and estimate the g-band magnitude from the corrected $G$,
$B_P$ and $B_P$ magnitudes following \cite{Riello2021}. The
reported scatter for these estimates is $0.075$~mag.
We then convert to the estimated counts in an image using
the zero point $Z$ of the ASAS-SN reference image and the mean
transparency correction $t$ between fluxes on the reference image and
the target image determined by ISIS (\citealt{Alard1998}, \citealt{Alard2000}) 
as part of the standard ASAS-SN image subtractions (the values in {\tt sum\_kernel}).
This gives predicted counts of
\begin{equation}
        \log N_s = - 0.4 (g-Z) - \log t.
\end{equation}

The ASAS-SN point spread function has a FWHM of roughly
16~arcsec. At this resolution, the flux in any ASAS-SN
photometric aperture can be a blend of the fluxes from 
multiple stars.  This is not important for the brightest
stars, but becomes increasingly important for fainter stars,
particularly at lower Galactic latitudes.   To mitigate this,
we found all $G<20$~mag Gaia DR3 stars within 1~arcmin 
of each target star.  We computed a G-band flux correction $f_c$
factor to go from the flux of the target star to the total
flux of all stars within $8$~arcsec. Assuming that the network 
will ``learn'' about background subtraction, we need to correct
this contaminating flux for the median stellar background flux
contribution, $b_c$, to the signal region.  Hence, we rescale
the prediction for the number of counts as 
\begin{equation}
       N_s \rightarrow N_s (1+f_c) - b_c.
\end{equation}

When we added bias subtraction to the ASAS-SN pipelines, we started to damage the cores of
the images
of saturated stars, a problem that remains to be fixed.  Such stars are
identifiable by the presence of pixels that are exactly unity because the final
consequence of the problem is that the modified ASAS pipeline declares them to be bad pixels
and resets them to unity.  Postage stamps with more than four such pixels are rejected for training and validation.

\section{Methods} \label{sec:methods}

\begin{figure}[h!]
    \centering
    \includegraphics[width=0.5\linewidth]{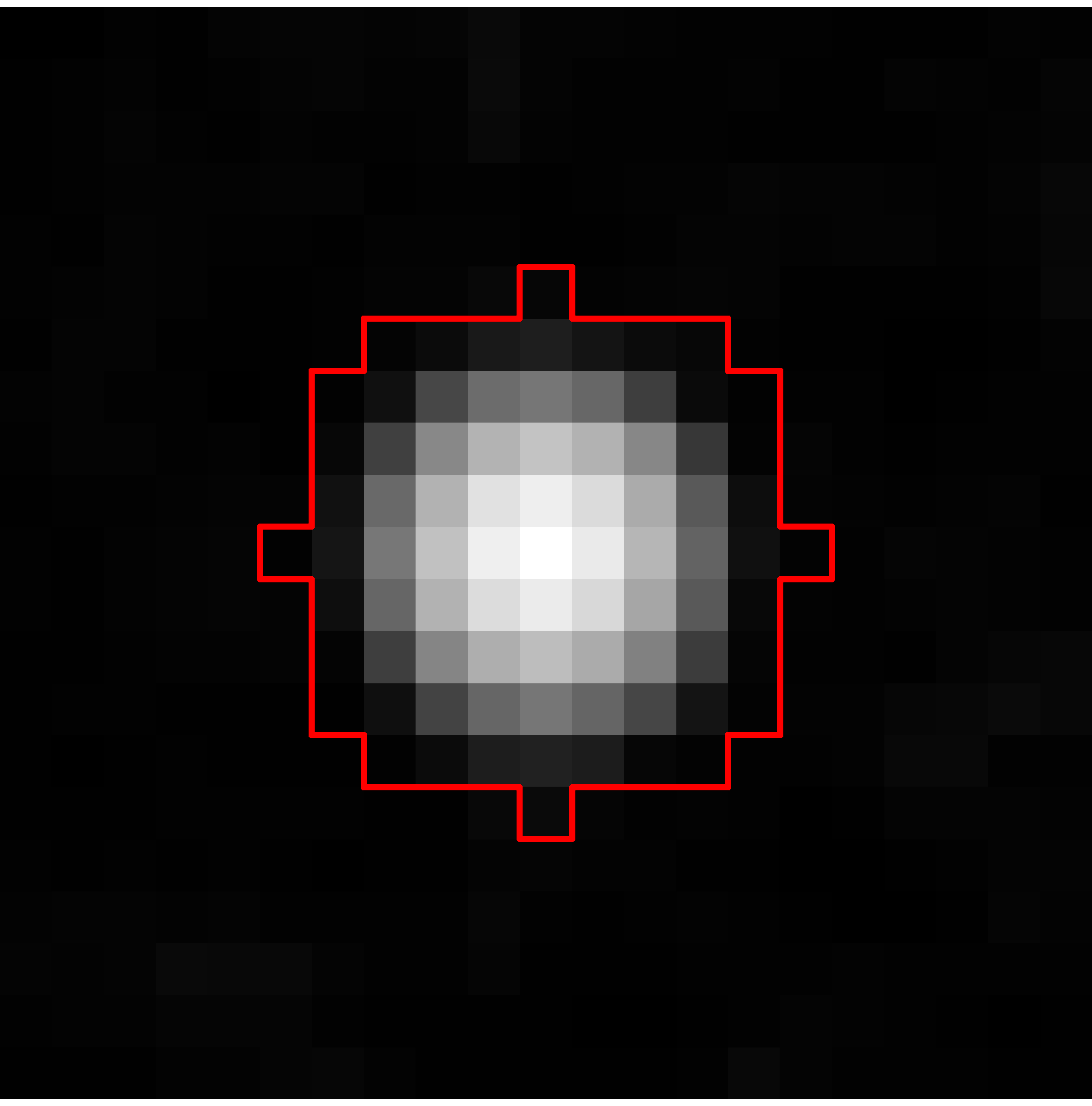}
    \caption{A log scale image of the postage stamp of a 99.9th percentile bright star with a red
      contour drawn to show which pixels are used in our model.}
    \label{fig:postage_stamp}
\end{figure}

In our dataset, the target star never covered more than a few pixels in any direction, even at the highest brightness levels.
As a pre-processing step, we reduced the analyzed region of each postage stamp to a 5 pixel radius circle around the center.
This is to help focus the model training on the relevant region and to give less context for a model to memorize.
A sampling of the brightest stars found that a 5 pixel radius would capture most visible detail with a $\sim$1-2 pixel margin, as shown in Fig.~\ref{fig:postage_stamp}.

As an easy way to expand the training set, we also considered reflections of the postage
stamps, adding three new postage stamps flipped horizontally, vertically, or both.  This 
grows the training set from $\sim 332$~thousand stamps to $\sim 1.33$~million stamps. To
the extent that the ASAS-SN point spread function is not circularly symmetric, this is
like moving the postage stamp to the corresponding reflection of its position on the 
detector.  We expected this to be a safe way of trivially multiplying the training set.
We tracked the results for both the original and permuted orientations and
were unable to find significant differences in the results.

Neural networks operate on the data turned into ``tensor data'' which is simply
a one dimensional array.  We can also append metadata to the end of the training data,
where we experimented with adding the reference image zeropoint $Z$ and the estimated
sky brightness of the individual image $S$. We discuss
other possible metadata that could be used in the discussion.  

The overall training
set is then divided into training (60\%), testing (20\%) and validation sets (20\%).
We were concerned about the potential for the network to memorize the expected 
properties of a particular star based on the pattern of surrounding background stars.
To make sure this was not an issue while validating the results, all images
of a particular star were placed in the same subset.

\section{Results} \label{sec:results}

\begin{figure*}
    \centering
    \includegraphics[width=0.9\linewidth]{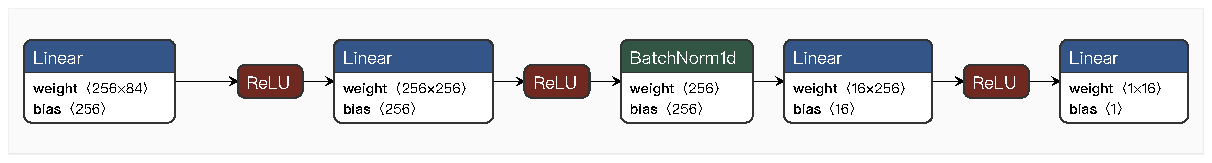}
    \caption{Final Model Architecture}
    \label{fig:model-arch}
\end{figure*}

We carried out our analysis using PyTorch~(\citealt{Paszke_PyTorch_An_Imperative_2019}), optimizing 
the mean squared error (MSE) loss function between the
input estimate of $\log N_s$ and the output estimate from the network. 
We began by experimenting with different MLP architectures.
We started with 2-4 smaller 64-wide layers, but found that increasing 
the width led to faster convergence. We ultimately
settled on 256-wide layers as 
increasing past this led to minimal improvements.
The final architecture is shown in Fig.~\ref{fig:model-arch}.

After selecting the architecture, 
we evaluated PyTorch's Adam, AdamW, Adadelta, Adagrad, Adamax, RMSProp, and SDG optimizers.
We trained each model to 150 epochs with each optimizer multiple times.
The Adam optimizer had the lowest losses, with AdamW coming second.
The other optimizers were less reliable for this task, with some never converging.
We used the Adam optimizer of \cite{kingma_adam_2017} for our final models, with a learning 
rate of 0.0001 and a batch size of 128.  We trained $24$ models using this architecture
including the hyper-parameters in the training and selected the one with the best MSE for
the validation data.

\begin{figure}
    \centering
    \includegraphics[width=0.95\linewidth]{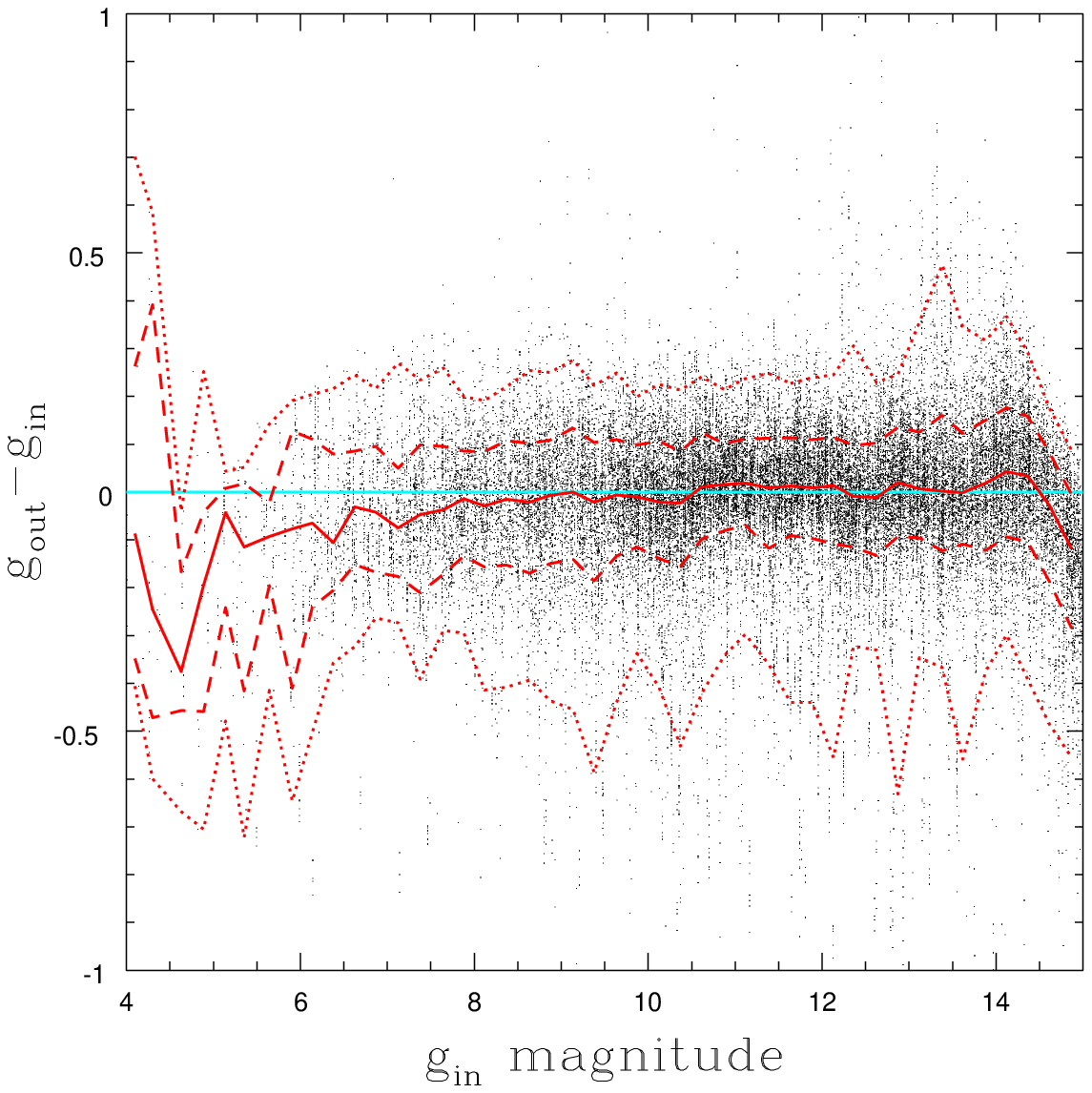}
    \caption{The differences between the input ($g_{in}$) and output
      ($g_{out}$) magnitudes for the verification stars.  The red
     curves show the median (solid), 68\% (dashed) and 95\% (dotted)
      ranges of the differences
     in bins of $0.25$~mag. A horizontal cyan line is included
      where the difference is zero.}
    \label{fig:model-error}
\end{figure}

The performance of the fiducial MLP model for the verification data set is shown in Fig.~\ref{fig:model-error}.
The results show little bias over 10 magnitudes (a factor of $10^4$ in flux) from $g \simeq 4$
mag to $g=14.5$~mag.  The median difference is $-0.004$~mag, with a clear bias appearing only for
the brightest and faintest stars.  The magnitude differences encompassing 68\% and 95\% of the stars are
$0.12$ and $0.34$~mag and the RMS dispersion is $0.168$~mag.  

We also considered several possible elaborations.
First, we considered whether ensemble results would perform better 
than the best model.  For example, if we defined the ensemble model
as the average of the results from the four individual models with the
lowest MSE values, the RMS residual did not improve. Second, we tried
averaging the results for the input image and its three permutations,
but again found no significant improvement. 
As noted earlier, many saturated stars have pixels reset to unity due to some 
unanticipated consequences of adding overscan corrections to the pipeline.  We
built an independent network trained only on objects with at least one pixel
set to unity to see if this would do better than simply mixing the two populations.
This model had an RMS residual of $0.360$~mag, far worse than the general analysis. 
We believe the small size of the resulting training set likely drives this.
From this point on, we focus on our fiducial MLP model.

We looked for correlations of the residuals with distance from the detector center and
Galactic latitude.  We examined the distance from the center because a source of a given flux produces $\sim 50\%$
fewer counts in the extreme corners of the detectors compared to the center due to 
vignetting.  We examined the Galactic latitude because the stellar density increases
rapidly for lower absolute latitudes, so the results depend more on correctly accounting
for crowding by other sources under the ASAS-SN PSF.  We found no significant correlation
of the residuals of both the saturated and unsaturated stars
with either the detector location or Galactic latitude.

\begin{figure}
    \centering
    \includegraphics[width=0.95\linewidth]{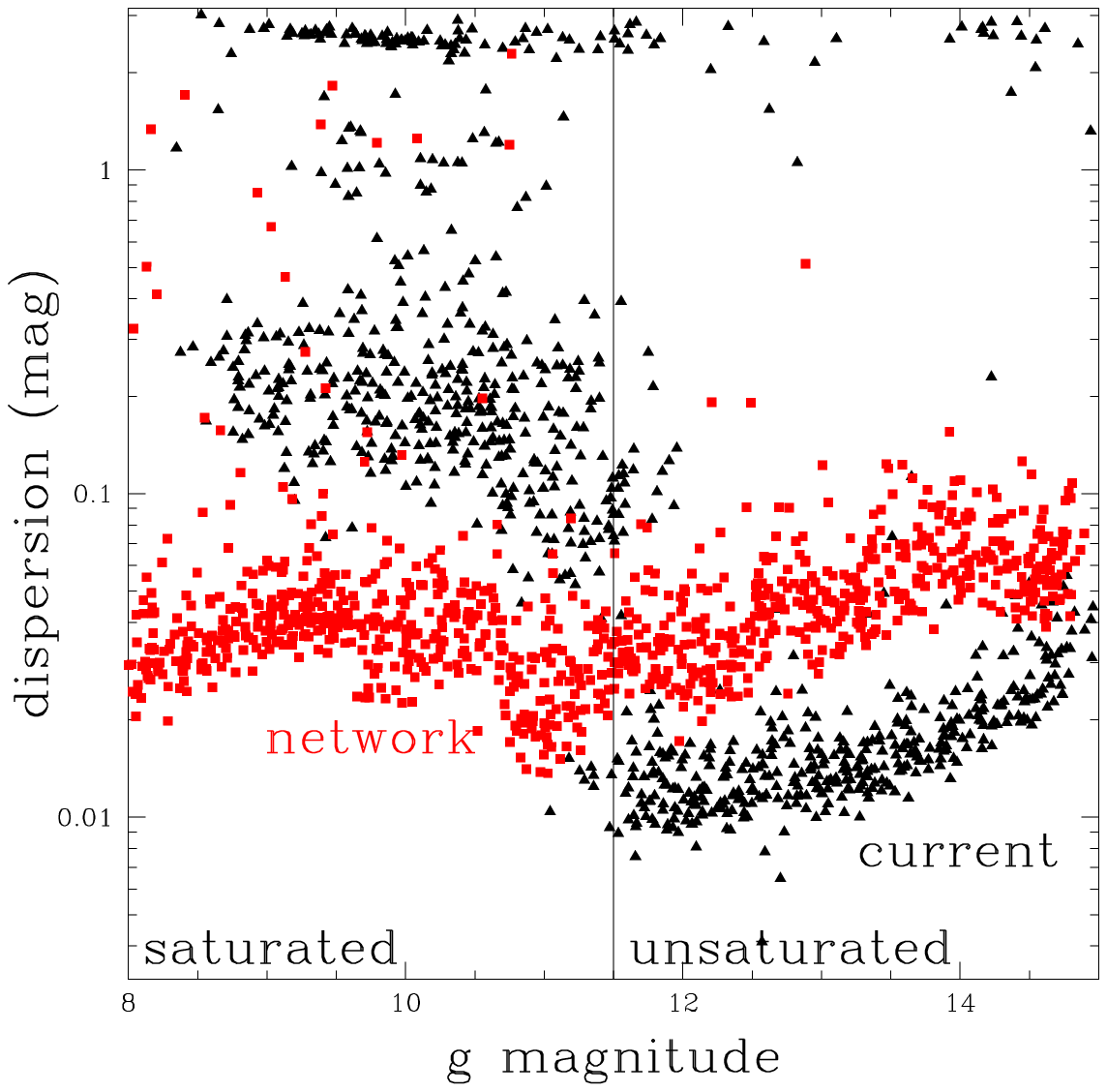}
    \caption{The light curve dispersions of approximately $10^3$ non-variable sources as a function of $g$
     magnitude analyzed using the current SP2 pipeline (black triangles) or by the neural network (red
      squares).  The dispersion is defined as one-half of the 15-85\% ($1\sigma$) range
      of the residuals about the median. The saturated (unsaturated) magnitude range is to
      the left (right) of the vertical line.}
    \label{fig:both}
\end{figure}

\begin{figure*}
    \centering
    \includegraphics[width=0.85\linewidth]{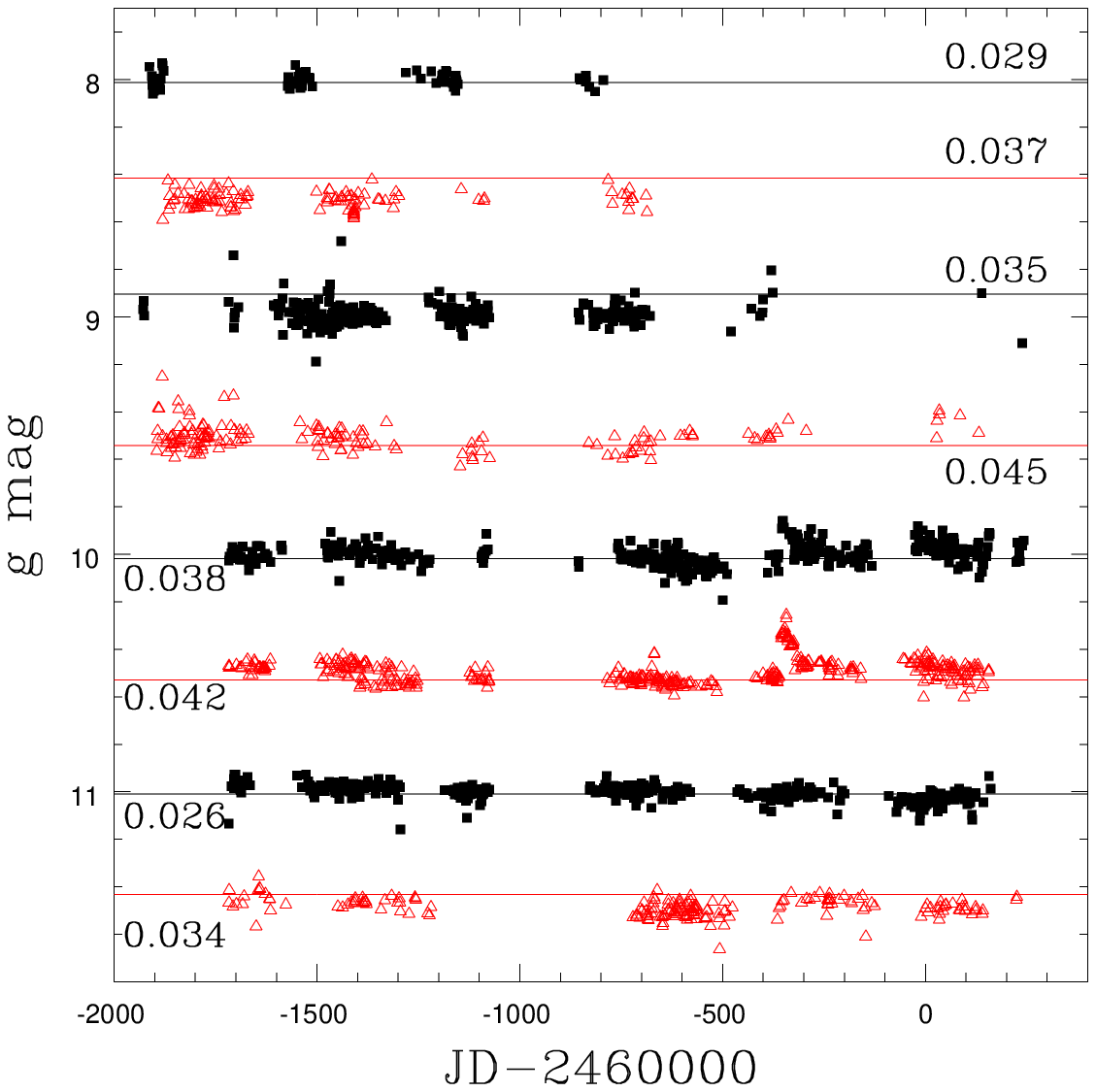}
    \caption{Light curves of the  
    stars in Fig.~\ref{fig:both} closest to $g=8$~mag to
    $11.5$~mag in steps of $0.5$~mag.  They
    were also required to have 68\% distribution
    widths less than $0.053$~mag -- 85\% of 
    $g<11.5$~mag stars have smaller dispersions.
    The horizontal lines are the Gaia-estimated
    g magnitudes. The number gives the dispersion
    estimated from the 15-85\% ($1\sigma$) range
    of the points about their median.
      }
    \label{fig:light-curve}
\end{figure*}

\begin{figure*}
    \centering
    \includegraphics[width=0.95\linewidth]{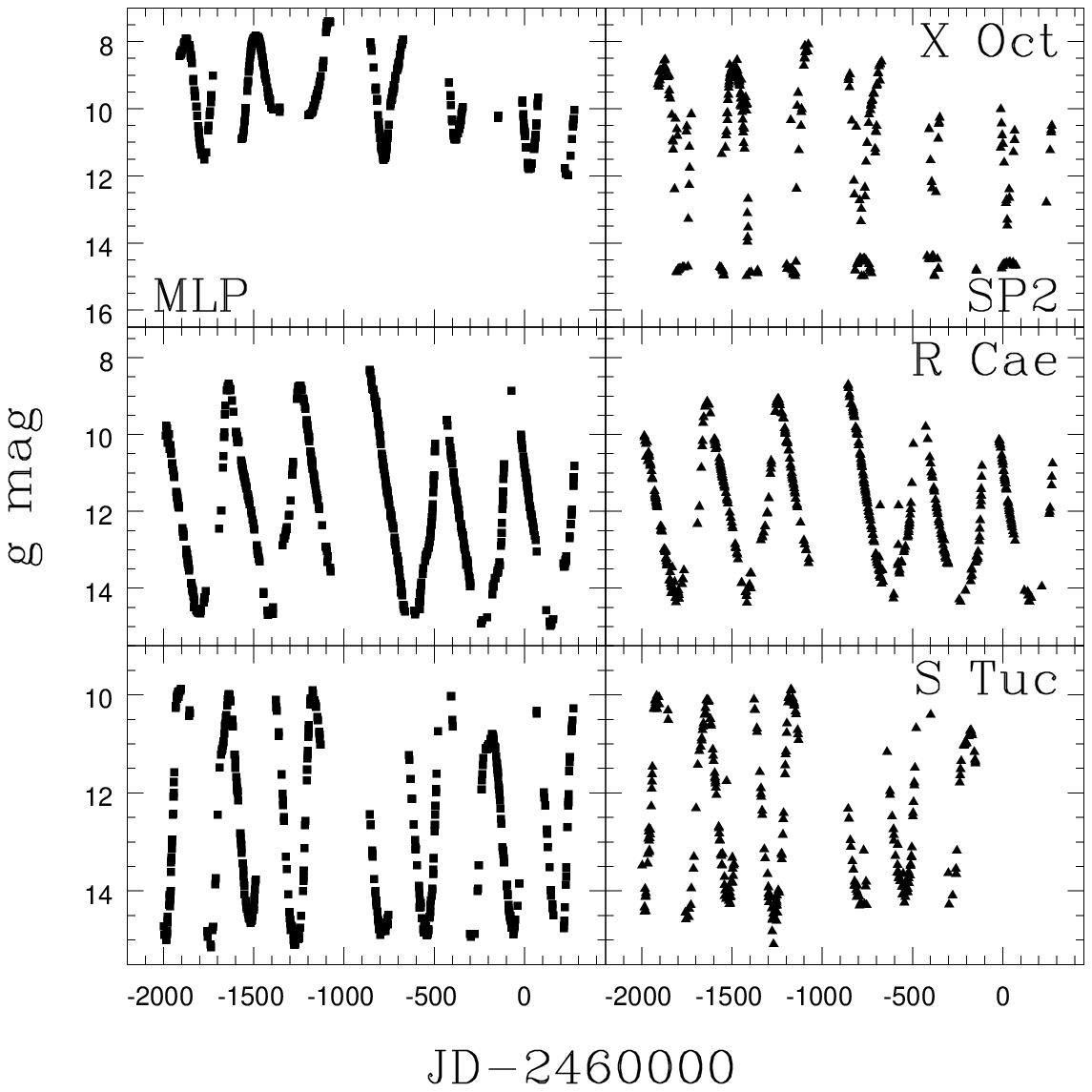}
    \caption{Light curves of Mira variables from the MLP analysis
    (left) and Sky Patrol V2.0 (right, SP2). While the saturated
     star corrections of the ASAS-SN pipeline produce surprisingly
      good results in many cases, the MLP pipeline results are 
      smoother and have fewer outliers like the ones prominently
       seen for the SP2 light curve of X~Oct.  X~Oct is also the
         example which is most 
       saturated at peak brightness. The vertical scales are
       the same for both the MLP and SP2 light curves.
       }
    \label{fig:mira}
\end{figure*}

\begin{figure*}
    \centering
    \includegraphics[width=0.95\linewidth]{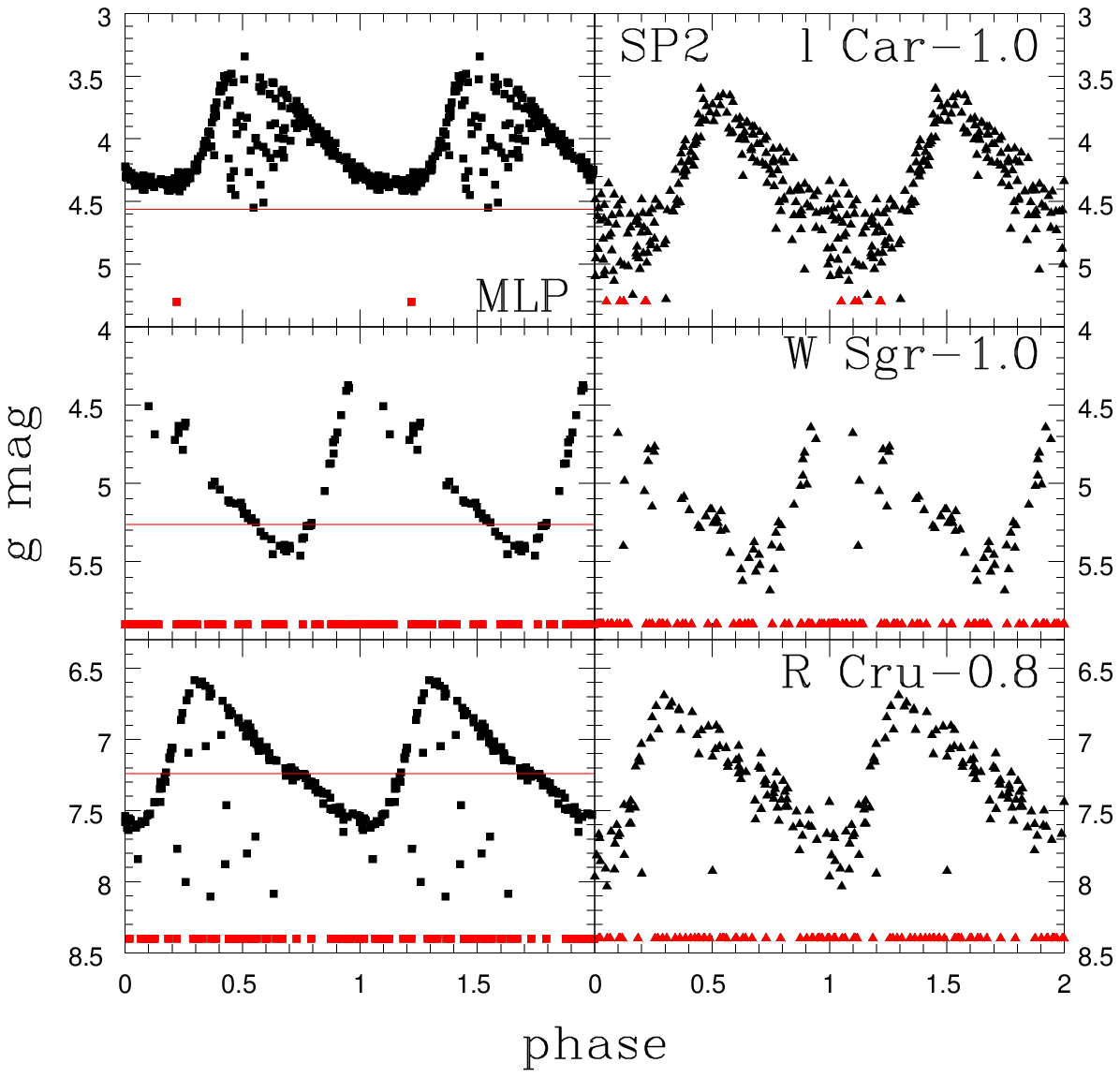}
    \caption{Light curves of very saturated
     Cepheid variables from the MLP analysis
    (left) and Sky Patrol V2.0 (right, SP2).  The
     SP2 light curves are shifted to brighter magnitudes
     by the amount next to the variable name.
    The red points
      lying in a line along the bottom are outliers fainter
    than the minimum magnitudes of the panels.  The horizontal
      red line is the g magnitude predicted from the mean Gaia
      magnitudes, which should be relatively close to the true
       mean magnitude.  Both pipelines produce many faint outliers
      for W~Sgr and R~Cru,
       but the MLP light curves are arguably cleaner except for $\ell$~Car at
       peak.
      The vertical scales are the same for both the MLP and SP2
       light curves.
       }
    \label{fig:ceph}
\end{figure*}

\begin{figure*}
    \centering
    \includegraphics[width=0.95\linewidth]{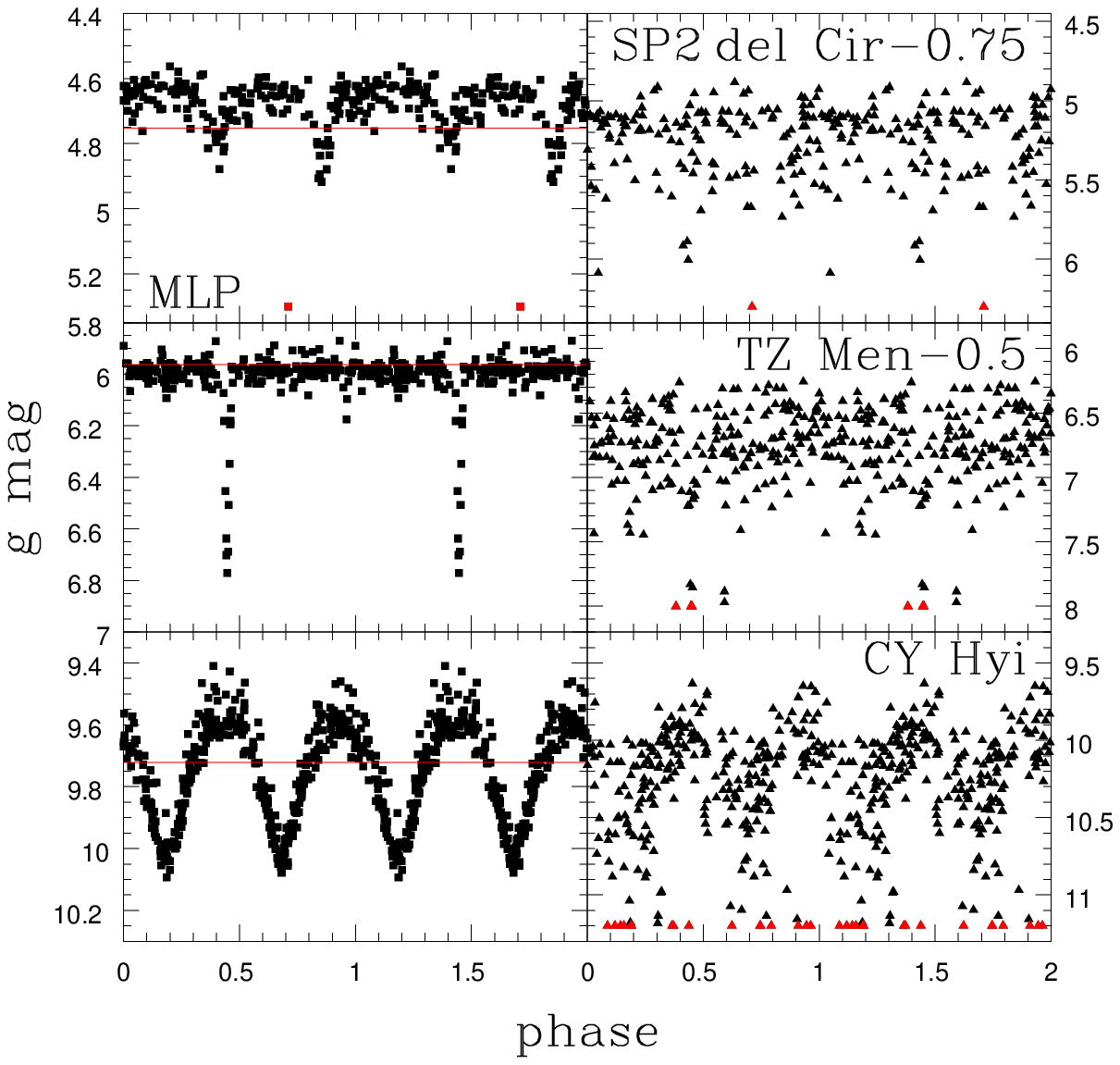}
    \caption{Light curves of eclipsing binaries from the MLP analysis
    (left) and Sky Patrol V2.0 (right, SP2).  The
     SP2 light curves are shifted to brighter magnitudes
     by the amount next to the variable name. The bright
      magnitude limit is the same for the MLP and SP2 panels,
      but the magnitude range is generally much larger for the
      SP2 panels. The red points
      lying in a line along the bottom are outliers fainter
    than the minimum magnitudes of the panels.  The horizontal
      red line is the g magnitude predicted from the mean Gaia
      magnitudes.
       }
    \label{fig:eb1}
\end{figure*}

\begin{figure*}
    \centering
    \includegraphics[width=0.95\linewidth]{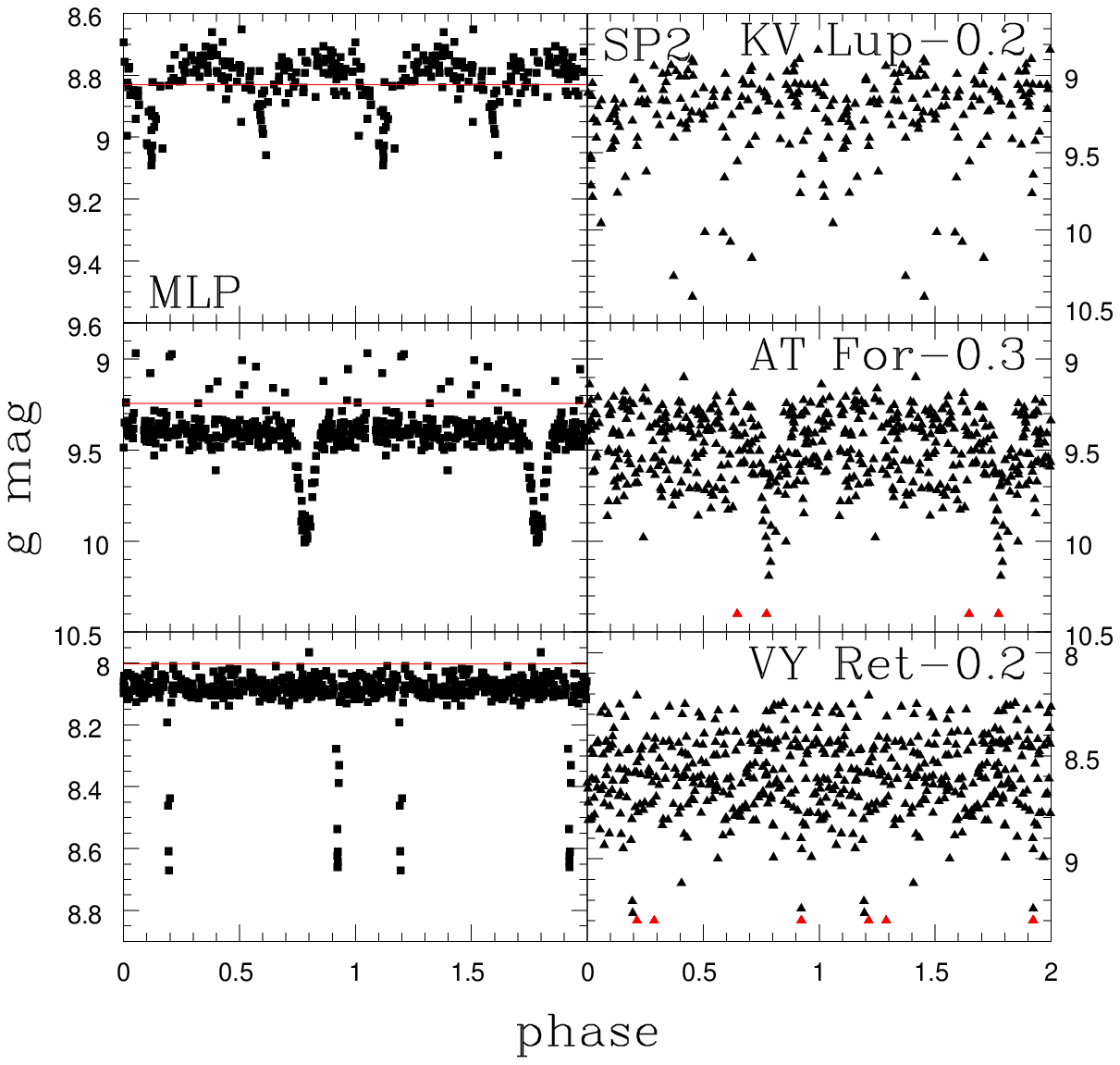}
    \caption{Light curves of eclipsing binaries from the MLP analysis
    (left) and Sky Patrol V2.0 (right, SP2).  The
     SP2 light curves are shifted to brighter magnitudes
     by the amount next to the variable name. The bright
      magnitude limit is the same for the MLP and SP2 panels,
      but the magnitude range is generally much larger for the
      SP2 panels. The red points
      lying in a line along the bottom are outliers fainter
    than the minimum magnitudes of the panels.  The horizontal
      red line is the g magnitude predicted from the mean Gaia
      magnitudes. VY~Ret has a highly eccentric orbit leading to
       the asymmetric eclipse phases.
       }
    \label{fig:eb2}
\end{figure*}

\begin{figure*}
    \centering
    \includegraphics[width=0.95\linewidth]{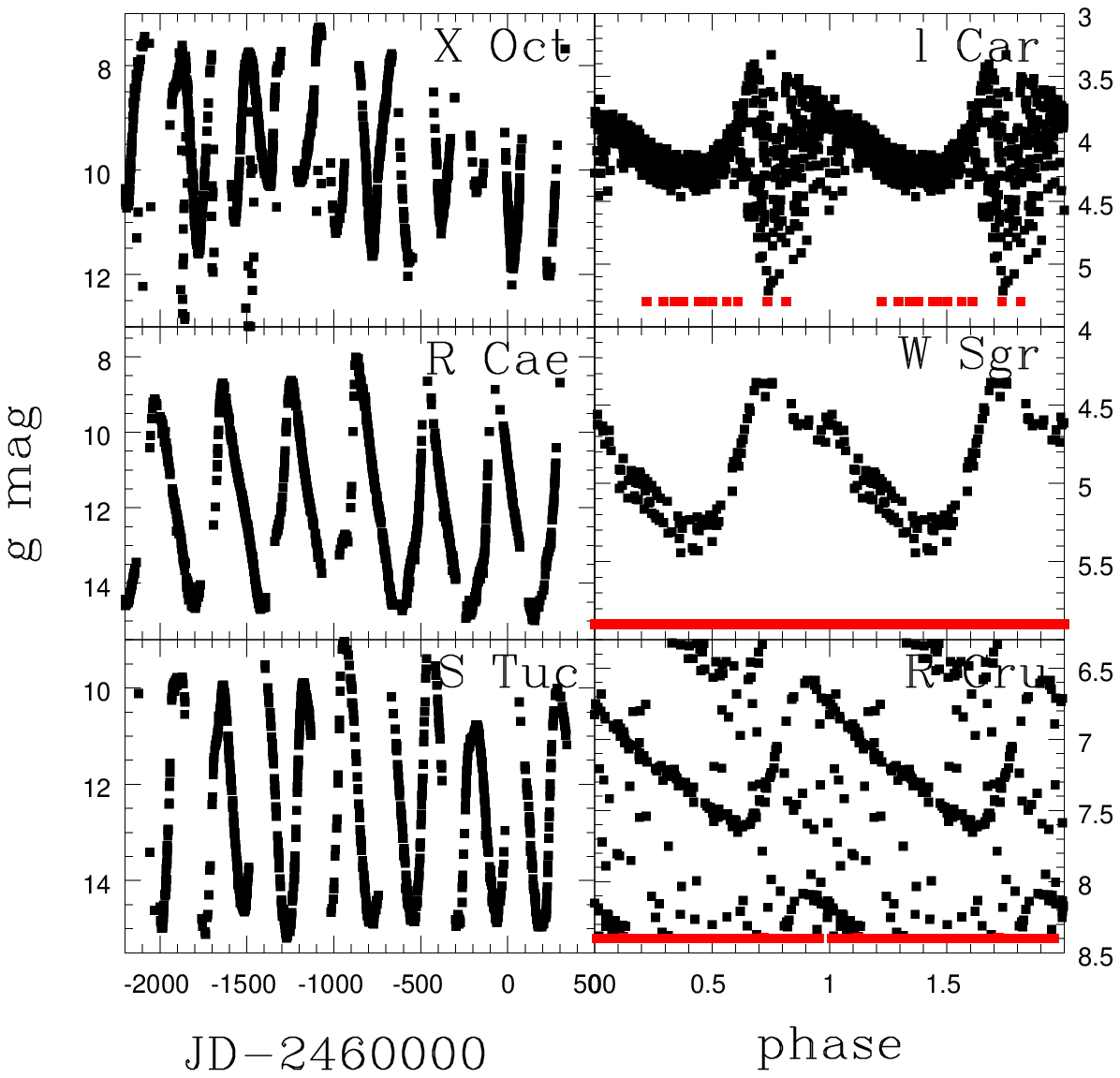}
    \caption{MLP light curves of the Miras and Cepheids
      from Figs.~\protect\ref{fig:mira} and \protect\ref{fig:ceph}
      (black squares) as compared to their MLP light curves using both V and
       g band data from all the ASAS-SN cameras observing each star.  Outliers are
       marked at the bottom in red.
       }
    \label{fig:came1}
\end{figure*}

\begin{figure*}
    \centering
    \includegraphics[width=0.95\linewidth]{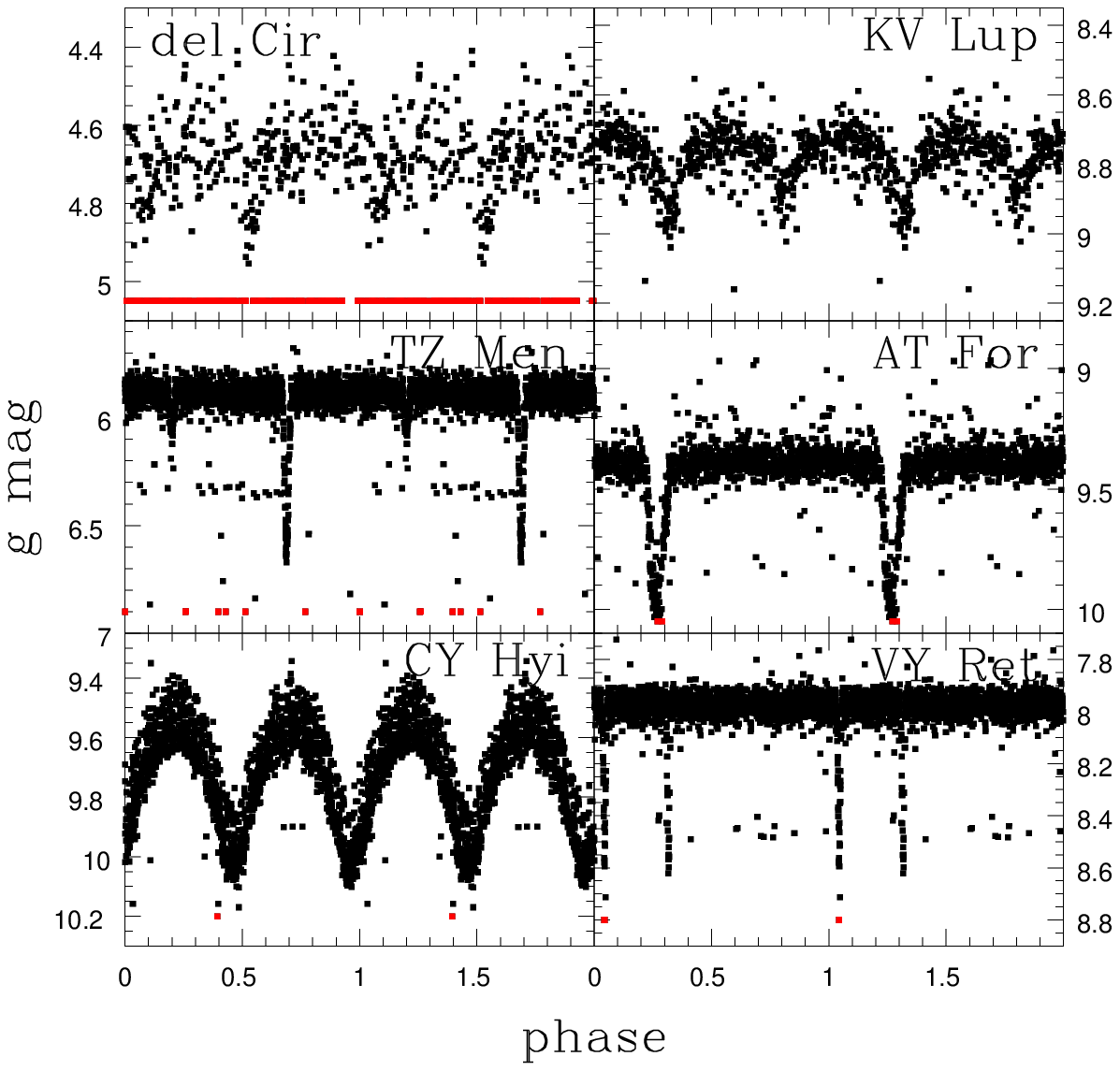}
    \caption{MLP light curves of the eclipsing binaries
      from Figs.~\protect\ref{fig:eb1} and \protect\ref{fig:eb2}
      (black squares) as compared to their light curves using both V and
       g band data from all the ASAS-SN cameras observing each star.  Outliers are
       marked at the bottom in red.
       }
    \label{fig:came2}
\end{figure*}

Fig.~\ref{fig:model-error} is not really a test of our primary goal, which is to
produce improved light curves rather than improved absolute photometry over the current
ASAS-SN pipelines as implemented in Sky Patrol v2.0 (SP2, \citealt{Hart2023}). 
ASAS-SN epochs are generally comprised of three 90 second exposures.
SP2 does aperture photometry on the co-added, subtracted images obtained for each
epoch and then adds the flux of the source in the reference image.   
The MLP analyzes individual images
since we are limited by systematic problems not photon statistics for the saturated
stars.  For the final MLP light curves we used the median of the results for the 
individual epoch images (the average if only two images, and the value if there is just one image).
This procedure noticeably reduces the light curve dispersions.

To test the performance on light curves, we randomly selected
1000 (Gaia) non-varying sources uniformly in magnitude over $8 < g < 14.5$~mag
and extracted their light curves using both the fiducial MLP model and SP2.  
We again limited the analysis to good
images and kept only images analyzed by both pipelines.  
Fig.~\ref{fig:both} compares the dispersions of the MLP and SP2 light curves with at least
50 points as a function of magnitude. The dispersions are defined by one-half of the 
15-85\% ($1\sigma$) range about the light curve median. The standard
pipeline has obvious problems for saturated ($g<11.5$) images, with a large jump in
the scatter from a median of $0.016$~mag for $g>11.5$ to a median of $0.22$~mag for
brighter sources.  The band of sources with scatters above one magnitude are the ones
most badly affected by the effects of the bias subtraction problem.  The performance of
the MLP is fairly uniform with a median scatter of $0.037$~mag for the brighter sources and
$0.048$ for the fainter sources.  A few of the high scatter outliers are high proper
motion stars, but most are cases where the bleed trail correction algorithm has put 
the flux from the bleed trail in the wrong location.  An obvious issue for future
work is to explore why the MLP has three times the scatter of the standard pipeline
for the unsaturated stars. 
Fig.~\ref{fig:light-curve} shows eight examples of these MLP light curves. They
were selected to have 15-85\% ($1\sigma$) ranges about their medians of $<0.053$~mag,
which is true for 85\% of the saturated stars (see Fig.~\ref{fig:both}), but were
otherwise just chosen to be the closest star to $8.0$, $8.5$ $\cdots$ $11.5$~mag.

Figs.~\ref{fig:mira}, \ref{fig:ceph}, \ref{fig:eb1} and \ref{fig:eb2} compare the
MLP light curves of Mira, Cepheid and eclipsing binaries to those from SP2.  
These variables were not selected to make
the MLP results look good, but simply to have interesting magnitudes and amplitudes
for illustrating results. The Miras illustrate the ability of the MLP analysis to 
smoothly track the brightness of variable stars from well below the saturation
limit near $g\simeq 11.25$~mag to significantly above it.  While the SP2 light
curves are frequently good, they start to produce significant outliers at the
brightest magnitudes, particularly for X~Oct.  The examples of classical Cepheids
include the extremely bright $\ell$~Car, the still naked eye visible W~Sgr and
the slightly fainter R~Cru.  The MLP $\ell$ Car light curves are producing 
significant numbers of outliers near peak, while the SP2 light curves struggle
near minimum. The W~Sgr and R~Cru light curves have significant numbers of faint
outliers, but the MLP produces generally smoother light curves if we ignore the outliers.  The
MLP magnitudes are also closer to the  g~magnitudes predicted from the
Gaia mean magnitudes. Note that since they are mean magnitudes, they should
be relatively immune to any variability.
The SP2 light curves are 
significantly fainter than the predictions and are shifted brighter by $\sim 1$~mag
in order to use the same magnitude ranges for both the MLP and SP2 light curves.

The driver of these problems for the bright Cepheids are two failure modes in the
pipeline saturated star correction.  One failure mode is to place
the flux from the bleed trails at the wrong location, usually a relatively
bright (but not nearly as bright) star lying nearby on the bleed trail.  The
second, which we do not understand, but appears to be the dominant failure for W~Sgr,
is to have not placed the missing flux anywhere.  The saturated target appears to
have been treated as part of the bleed trail and filled in by interpolation across
the estimated trail.  

The eclipsing binary comparisons provide the most dramatic illustration of the 
improvements from the MLP approach - the eclipses are clearly visible in the MLP light curves and
essentially invisible in the SP2 light curves.  Even 
CY~Hyi, with its fairly high amplitude and continuous variability with phase,
has a barely discernable variability pattern in the SP2 light curve.
 The shallow eclipses of del~Cir and KV~Lup are
clear for the MLP light curves, as are the very narrow eclipses of TZ~Men and
VY~Ret.  The mean MLP magnitudes are close to the predicted means, while the SP2
means are again significantly fainter than predicted. There are outliers in the
MLP light curves, but most could probably be eliminated with some variant
of sigma clipping.

The MLP was trained using data from only a single camera (``bi'' on
the Paczynski mount), while ASAS-SN
is presently comprised of 20 cameras. ASAS-SN also initially used
V band, while the training data was all g band photometry.  As an
experiment, we simply analyzed the data from all cameras and
both filters for the stars in Figs.~\ref{fig:mira},
\ref{fig:ceph}, \ref{fig:eb1} and \ref{fig:eb2}. We ``inter-calibrated''
the data from all the sources using a Damped Random Walk (DRW) Gaussian
process for interpolation and solving for the best light curve means for
the individual cameras and filters (these are a form of the ``linear
parameters'' discussed in \cite{Kozlowski2010} for DRW models
of quasar light curves).  We found we could
reduce the effects of faint outliers by dropping the faintest 10\% of
the light curves.  Each light curve was then offset by the difference
between the median light curve mean and the mean for the individual
camera and filter.  This does assume that the variability amplitudes
and light curves shapes are identical for the two filters, but it also
the most extreme test for seeing how crucial it will be to individually
train by camera and filter.

Figs.~\ref{fig:came1} and \ref{fig:came2} show the results. They
are generally very good, suggesting that the current network can simply
be applied generally and that we could train one MLP using data from all cameras
and filters. There are outliers such as those seen for X~Oct, 
and all the data are struggling with the bright Cepheids from 
Fig.~\ref{fig:came1}.  The biggest failure is for R Cru, where
the huge number of outliers has confused the inter-calibration
procedure.  This is seen to a lesser extent for W~Sgr, where too
few faint outliers were rejected for the good parts of the light
curves to be perfectly lined up.  But for R Cae, S Tuc and the
eclipsing binaries in Fig.~\ref{fig:came2}, the results are 
extremely good.

\section{Conclusions and Future Work} \label{sec:conclusion}

We developed a multilevel perceptron (MLP) neural network for the accurate photometry of saturated stars.
The network was trained using predictions for the counts produced by the target star
in a sample of $\sim 332$ thousand postage stamp images of stars roughly uniformly 
spread in magnitude from $g \sim 3 $ to $15$~mag. For the verification sample, the
median magnitude differences are well within the 68\% scatter of the differences
($0.12$~mag) except for the very brightest and faintest stars (see Fig.~\ref{fig:model-error}). 
More importantly, the
typical scatter in the light curve of a saturated star is only $0.037$~mag (half
the 15 to 85 percentile range about the mean) compared to $0.22$~mag for the SP2
light curves (see Fig.~\ref{fig:both}).  The MLP light curves of many bright
variable stars are dramatically better than their SP2 light curves (see Figs.~\ref{fig:mira} to \ref{fig:eb2}) although the performance can sometimes be poor, as illustrated
by the naked eye Cepheids in Fig.~\ref{fig:ceph} (see below).  Although the network
was trained using data from only one camera, it appears to work equally well when
tested using data from all cameras and both V and g band data when combined
using a Gaussian process to intercalibrate the light curve means
(see Figs.~\ref{fig:came1}, \ref{fig:came2}).  For some systems, this calibration
procedure and/or outlier rejection would have to be done more carefully, but the
results are otherwise very encouraging.

We analyzed the reduced ASAS-SN images interpolated to the frame of the reference image.
Using the interpolated images meant that the pixel location of the target stars was 
fixed and could be accurately determined from the well-verified astrometry of
the reference images (2~arcsec errors relative to Tycho stars at worst, compared to a
16~arcsec FWHM).  We do not see any evidence that the interpolation causes problems.

Using the as-reduced images creates the problems seen for the bright
Cepheids because of two
issues.  The first is that an interplay between adding bias corrections and
the inheritance from the ASAS pipeline of reading only integer fits files can lead to
damaging saturated images because of integer overflows. This manifests as pixels in the
star being flagged as bad and given a pixel value of unity.  This obviously causes 
severe problems for the standard aperture photometry pipeline.  The MLP pipeline 
largely manages to correctly interpret the flagged pixels and recover a good 
estimate of the true flux.  While we understand the origin of the problem, a fix is not 
trivial and has yet to be implemented.  
We believe the performance of a MLP retrained after the problem is fixed should be
improved.

The second problem comes from mistakes made by the saturated star corrections 
inherited from the ASAS pipeline.  As noted earlier, the pipeline tries to collect
the flux from the bleed trail and add it as a Gaussian at the location of the
saturated star with the FWHM of the data.  This generally works very well, but
the flux is sometimes assigned to the wrong location and sometimes seems not 
to be assigned to any location.  We see this problem here for the very brightest
examples we show (Figs.~\ref{fig:ceph}, \ref{fig:came1}, \ref{fig:came2}).
Determining the origin of these problems in the pipeline is beyond our present
scope.

We suspect that our approach would have worked well and avoided this problem using data
without these attempts to correct saturated stars.  Testing this is, however,
an involved process.  The original raw images are available, but they would have to
be reprocessed without the saturated star corrections (or at least the postage stamp
images needed for each star would have to be reprocessed).  The pixel positions of
stars would also now vary from image to image, so we would need to be fully 
confident of the astrometry of the individual images.  Checking this astrometry
has not been a priority because all the current ASAS-SN results depend only on
the astrometry of the reference image for a field and not on the astrometry of
the individual images.

One likely area where our results could be improved is in the estimation of the
predicted counts for the training set.  As a reminder, we predicted SDSS g magnitudes
from Gaia DR3 magnitudes, converted them to counts using the zero point of the ASAS-SN
reference image and the image subtraction estimate of the transparency difference between
the images defining the zero point and the current image, and attempted to correct for
the flux from other stars within the large ASAS-SN PSF.  The \cite{Riello2021} transformation
from Gaia magnitudes to g band is not great, with a reported scatter of $0.075$~mag. This
is, however, not large enough to be a major contributor to the scatter seen for the
verification data.  Nonetheless, it might be better to have used ATLAS REFCAT (\citealt{Tonry2018refcat}) 
since it was built in part to systematically estimate $g$ band magnitudes.

The crowding corrections were computed based only on
the Gaia G band magnitudes, and they should probably have been based on the estimated
g band magnitudes.  The thought was that these corrections should mostly be small, so 
modest fractional errors in them would not be important.  Retrospectively, this may be
need to be treated more carefully for the non-saturated stars.  There might also be
issues from the interplay between the estimated crowding corrections and the size
and geometry of the postage stamp images (Fig.~\ref{fig:postage_stamp}).
For the saturated stars, which are the primary target of the project, this is less of
an issue simply because the stars are so bright and the standard pipeline performs so
poorly.  Moreover, the scatter seen for the unsaturated stars in the verification
data does not depend on Galactic latitude, which suggests that problems with
confusion cannot be a dominant driver of the scatter.  

If a primary driver of the scatter was simply the accuracy of the conversion from
estimated counts to magnitudes driven by zero point and transparency errors, then
we would expect the fluctuations in the light curves of constant sources lying in
the same field to be correlated.  If so, this would then lead to a natural approach
to improving the image calibrations by using the correlated variability to estimate
an image calibration correction.  We inspected this question for the trial light curves 
of non-varying sources lying in the same field and found, unfortunately, no obvious
correlations.

The most important issues to be addressed are the problems created by the
ASAS-SN pipeline, fixing the damage done to the images of saturated stars and either understanding
the failures of the saturated star corrections or switching to using images without
the corrections.  We need to more extensively test how well the MLP works on multiple
cameras to determine if we can simply train a single MLP for all cameras, perhaps with
the camera identification as an additional item of metadata.  

This method is now available as an option on ASAS-SN Sky Patrol v1.0 (https://asas-sn.osu.edu/,
\citealt{Kochanek2017}) as the ``Saturated Stars (Machine Learning)'' photometry option.  As 
should be clear from the examples, the results become less predictable for stars
approaching naked eye visibility and brighter.  The present method supplies no
magnitude uncertainties, so these are simply reported as the typical scatter 
seen in Fig.~\ref{fig:both}.  Better error estimates of the relative errors
can be derived from the light curves, but this requires some information on the
nature of any variability (e.g., periodic or non-periodic), making it difficult to
estimate automatically. We
anticipate that the machine learning results will steadily improve, particularly once the reduction pipeline problems
for saturated stars are corrected. It is also considerably faster than the
most equivalent mode of Sky Patrol v1.0 (``Aperture Photometry''), which takes roughly
5 times longer to produce the light curves shown in Figs.~\ref{fig:mira}-\ref{fig:eb2}.

\begin{acknowledgements}
We thank Dustin Perzanowski of Ohio State's ASCTech
group for adding the machine learning option to 
Sky Patrol v1.0.
CSK is supported by NSF grants AST-2307385
and AST-1908570. ASAS-SN is funded in part by the Gordon and Betty Moore Foundation
through grants GBMF5490 and GBMF10501 to the Ohio State University, and also funded
in part by the Alfred P. Sloan Foundation grant G-2021-14192.
\end{acknowledgements}

\bibliography{sample631}{}
\bibliographystyle{aasjournal}



\end{document}